\documentclass[aps,prb,onecolumn,english,showpacs,superscriptaddress,amssymb,amsfonts]{revtex4}
\usepackage[T1]{fontenc}
\usepackage[latin9]{inputenc}
\usepackage{amsmath}
\usepackage{dsfont}
\usepackage{amstext}

\usepackage{tocvsec2}

\usepackage{amssymb}
\usepackage{amsbsy}
\usepackage{amsthm}
\usepackage{epsfig}
\usepackage{graphicx}
\usepackage{bbm}
\usepackage{hyperref}
\usepackage{color}
\usepackage{multirow}
\usepackage{pdfsync}
\def\beq{\begin{equation}}
\def\eeq{\end{equation}}

\newcommand{\Ref}[1]{Ref.~\onlinecite{#1}}

\newcommand{\bst}{{\mathcal{T}}}
\newcommand{\bsc}{{\mathcal{C}}}

\newcommand{\ie}{{\emph{i.e.~}}}
\makeatletter

\newcommand{\Rmnum}[1]{\expandafter\@slowromancap\romannumeral #1@}
\makeatother
\newcommand{\imth}{\hspace{1pt}\mathrm{i}\hspace{1pt}}

\newcommand{\eg}{{\emph{e.g.~}}}

\newcommand{\mbz}{{\mathbb{Z}}}
\newcommand{\bea}{\begin{eqnarray}}
\newcommand{\eea}{\end{eqnarray}}
\newcommand{\bpm}{\begin{pmatrix}}
\newcommand{\epm}{\end{pmatrix}}
\newcommand{\bal}{\begin{aligned}}
\newcommand{\eal}{\end{aligned}}

\newcommand{\dket}[1]{|{#1}\rangle}

\makeatother

\usepackage{babel}

\begin{document}
\title{Topological nature of step edge states on the surface of topological crystalline insulator Pb$_{0.7}$Sn$_{0.3}$Se}

\author{Davide Iaia}
\affiliation{Department of Physics and Frederick Seitz Materials Research Laboratory, University of Illinois Urbana-Champaign, Urbana, Illinois 61801, USA}
\author{Chang-Yan Wang}
\affiliation{Department of Physics, The Ohio State University, Columbus, OH 43210, USA}
\author{Yulia Maximenko}
\affiliation{Department of Physics and Frederick Seitz Materials Research Laboratory, University of Illinois Urbana-Champaign, Urbana, Illinois 61801, USA}
\author{Daniel Walkup}
\affiliation{National Institute of Standards and Technology, Gaithersburg, MD 20899, USA}
\author{R. Sankar}
\affiliation{Institute of Physics, Academia Sinica, Taipei R.O.C. 11529, Taiwan}
\author{Fangcheng Chou}
\affiliation{Center for Condensed Matter Sciences, National Taiwan University, Taipei 10617, Taiwan}
\author{Yuan-Ming Lu}
\affiliation{Department of Physics, The Ohio State University, Columbus, OH 43210, USA}
\author{Vidya Madhavan}
\affiliation{Department of Physics and Frederick Seitz Materials Research Laboratory, University of Illinois Urbana-Champaign, Urbana, Illinois 61801, USA}

\begin{abstract}
In addition to novel surface states, topological insulators can also exhibit robust gapless states at crystalline defects. Step edges constitute a class of common defects on the surface of crystals. In this work we establish the topological nature of one-dimensional (1D) bound states localized at step edges of the [001] surface of a topological crystalline insulator (TCI) Pb$_{0.7}$Sn$_{0.3}$Se, both theoretically and experimentally. We show that the topological stability of the step edge states arises from an emergent particle-hole symmetry of the surface low-energy physics, and demonstrate the experimental signatures of the particle-hole symmetry breaking. We also reveal the effects of an external magnetic field on the 1D bound states. Our work suggests the possibility of similar topological step edge modes in other topological materials with a rocks-salt structure.
\end{abstract}

\maketitle

\section{Introduction}

\newcommand{\didv}{$\text{d}I/\text{d}V$}

The discovery of topological insulators\cite{Hasan2010,Hasan2011,Qi2011} (TIs) has unearthed a large class of novel quantum materials, which host robust gapless surface excitations protected by various symmetries. In addition to two-dimensional (2D) surface states, certain topological materials can also host one-dimensional (1D) topological gapless modes at special crystalline defects such as lattice dislocations\cite{Ran2009,Juricic2012,Slager2014}. The topological nature of these 1D states offers them protection against back-scattering which can provide a dissipationless transport channel that might prove useful in device applications. Recently, 1D modes were discovered at odd step edges of a topological crystalline insulator (TCI)\cite{Sessi2016}, which were shown persist to high temperatures making them potentially attractive for applications. However, the topological character of these states, and hence their robustness against perturbations has not yet been established.

In this work, we elucidate the topological nature of the in-gap states localized at step edges on the surface of a topological crystalline insulator\cite{Hsieh2012,Ando2015} (TCI) Pb$_{1-x}$Sn$_x$Se. We use a combination of theory and scanning tunneling microscopy (STM) and spectroscopy (STS) to clarify the topological nature and classification of both odd and even step edge states.  Theoretically, by combining topological classification with microscopic calculations based on $k\cdot p$ theory\cite{Hsieh2012,Wang2013m,Serbyn2014,Rechcinski2018}, we reveal 1D flat bands localized at odd- and even-step edges, which we find are similar to those on the zigzag edge of graphene\cite{Nakada1996}. In particular, we show that the in-gap zero modes at the odd step edges form a Kramers pair and are topologically protected by an emergent particle-hole symmetry of the low-energy Hamiltonian.  We further predict that at even step edges, mixing of the two Kramers pairs is allowed and can lead to split peak features in density of states (DOS). These peaks can further split in a magnetic field. Our STS data confirm these predictions.  Our work therefore establishes the topological nature and stability of 1D step edge states in Pb$_{0.7}$Sn$_{0.3}$Se. Importantly, the general mechanism shown here can also apply to step edges in other rock-salt crystals with topological surface states.

\begin{figure}
\includegraphics[width=\columnwidth]{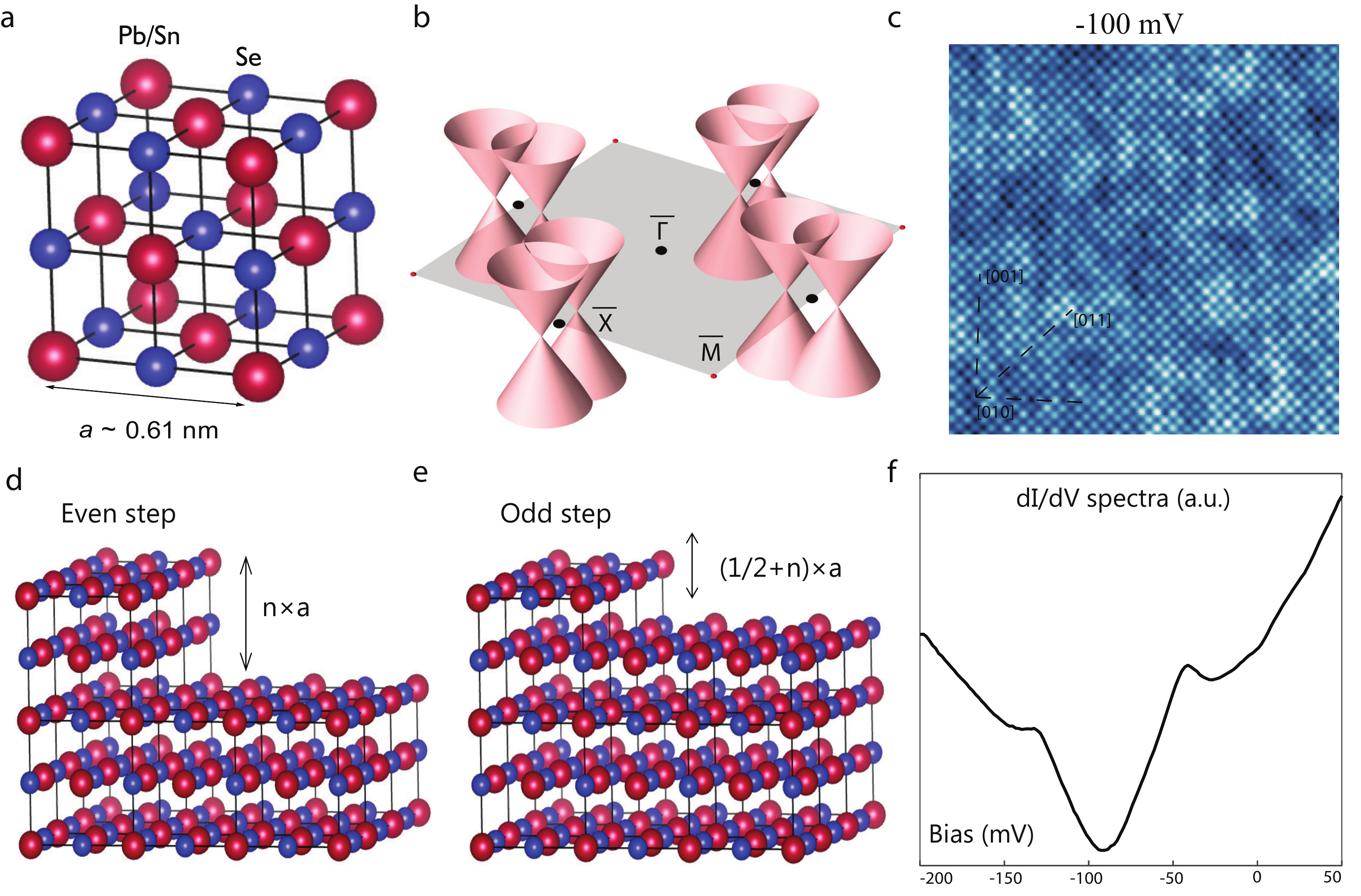}
\caption{Crystal structure and band structure. (a) Rock-salt crystal structure of Pb$_{0.7}$Sn$_{0.3}$Se. (b) Schematic band structure of the surface state. (c) $(15nm)^2$ STM topography at -100 mV, 200 pA and at 4 K. (d) Schematic representation of an even step. (e) Schematic representation of an odd step. (f) Typical averaged \didv~spectrum on the [001] surface of Pb$_{0.7}$Sn$_{0.3}$Se.}
\label{fig:1}
\end{figure}

\section{Results}

Pb$_{0.7}$Sn$_{0.3}$Se crystallizes in a rock-salt structure (Fig.\ref{fig:1} a) and can be easily cleaved to expose the [001] surface. Fig.\ref{fig:1}c shows a topographic image of this surface recorded at low bias. It is known\cite{Zeljkovic2015} that the low bias STM topography reveals only the Se sublattice, while high bias images contain information from both Se and the Pb/Sn sublattices. Correspondingly, we observe only the Se atoms in this image. The electronic and topological properties of Pb$_{1-x}$Sn$_x$Se have been widely investigated, theoretically\cite{Hsieh2012,Wang2013m,Safaei2013,Liu2013c} and experimentally\cite{Tanaka2012,Xu2012b,Neupane2015,Okada2013,Zeljkovic2014,Zeljkovic2015,Zeljkovic2015a,Walkup2018,Polley2018}. While it is not a strong topological insulator with a trivial $Z_2$ index, Pb$_{0.7}$Sn$_{0.3}$Se is a topological crystalline insulator characterized by a non-zero mirror Chern number\cite{Hsieh2012}. The [001] surface hosts four hybridized Dirac cones protected by mirror symmetry, symmetrically away from the $\bar X$  and $\bar Y$ points of the surface Brillouin zone (SBZ, see Fig.\ref{fig:1} b). The hybridization of the cones results in a change of the Fermi surface topology, known as the Lifshitz transition, when we move deep into the band gap. The Lifshitz transition is associated to a singularity in the density of states (DOS), known as the Van Hove singularity. Fig.\ref{fig:1}f shows a typical \didv~spectrum, which is proportional to the density of the states. The curve is V-shaped with a minimum at the Dirac point around $-90$ mV, and two peaks associated to the Van Hove singularity at $\sim-40$ mV and $\sim-130$ mV.

\begin{figure}
\includegraphics[width=0.8\columnwidth]{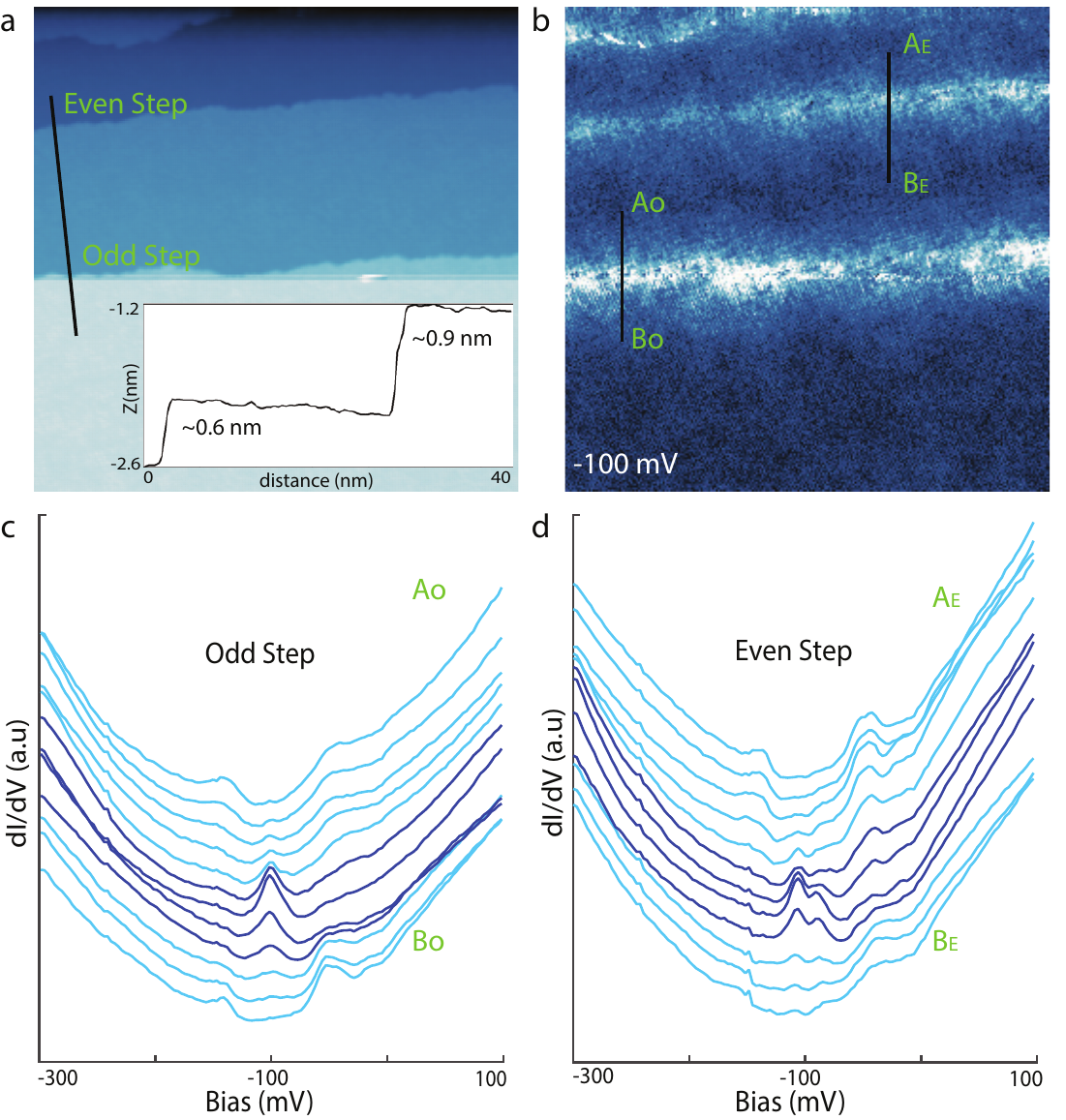}
\caption{Spectroscopy on even and odd steps. (a) $(60nm)^2$ topography image. The inset shows the height profile along the black line. (b) d$I$/d$V$ map map at -100 mV. (c) Spectroscopic linecut across the odd step. (d) Spectroscopic linecut across the even step.}
\label{fig:2}
\end{figure}

Since the unit cell consists of three atomic layers, two types of steps can be found in STM topography: even-steps whose height is an integer multiple of the lattice constant $a$ (Fig.\ref{fig:1}d); and odd-steps whose height is a half integer multiple of the lattice constant $a$ (Fig.\ref{fig:1}e). Our first task is to measure the local density of states near the even- and odd-step edges.  Fig.\ref{fig:2} a shows a topographic image with two step edges with the height profile measured across the step edges shown in the inset. From the heights we find two different type of step edges: an odd step of height $\frac32a\approx0.9$ nm and an even step of height $a\approx0.6$ nm.

A \didv~map recorded in the same area at energies close to the Dirac point (Fig.\ref{fig:2}b) shows a clear enhancement of density of states along the step edges. As first observed in \Ref{Sessi2016}, an enhancement of the DOS localized at odd steps is associated with the existence of robust one-dimensional electron channels connecting two Dirac cones. However the topological nature of these localized states has not yet been clarified. Moreover, unlike previous work, we also find a DOS enhancement at the even step edge. To obtain further information on these states we obtain several \didv~spectra (line cuts) across the steps. As Fig.\ref{fig:2} c shows, for the odd step a clear sharp peak is observable, as we cross the step edge. In accordance with the \didv~map, we also observe a peak-like feature in the DOS at the even step edge. The line shape of this feature however appears as a split peak (see Fig.\ref{fig:2} d). Our first goal is to theoretically understand the presence and characteristics of these step edge modes.

\begin{figure}
\includegraphics[width=1\columnwidth]{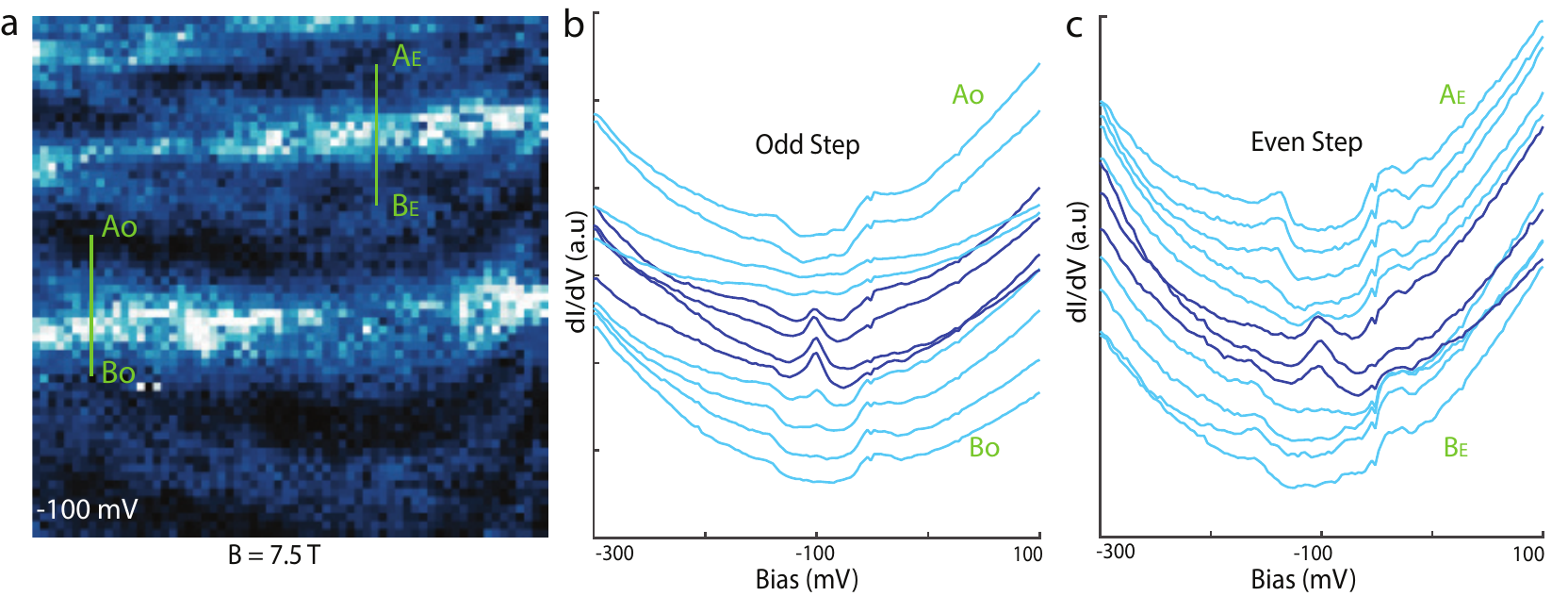}
\caption{Magnetic field data. (a) d$I$/d$V$ map at -100 mV. The area used for the map is the same as shown in Fig.2 a and b. (b) Line cut across the odd step edge. The line cut is in the same position (see  green line of Fig. 5 a) as in Fig. 2 c. (c) Line cut across the even step edge. The line cut is along the same position as the one shown in Fig. 2 d.}
\label{fig:3}
\end{figure}

To further characterize and distinguish between the odd- and even-step edge modes, we investigated their behavior under an external magnetic field. Fig.\ref{fig:3} a shows a d$I$/d$V$ map recorded with an out-of-plane magnetic field of 7.5T, in the same area used for Fig. 2a and b. Fig.\ref{fig:3} b and c are line cuts across the odd and even step, respectively, along the same position used for Fig.2 c and d. As it is evident from Fig.\ref{fig:3} d (also see Fig. 7 in supplement), the two peaks on the even step merge into one broad peak when we apply an external magnetic field.

 To theoretically reveal the topological index and stability of these  1D step edge modes we perform explicit calculations based on the $k\cdot p$ model of Pb$_{0.7}$Sn$_{0.3}$Se [001] surface states, as we discuss in detail in the next section.

\section{Discussion}

The low energy physics of [001] surface states of Pb$_{0.7}$Sn$_{0.3}$Se is described by the following $k\cdot p$ theory\cite{Wang2013m}
\bea
 h^{\bar X}_{\bf k}=-m\sigma_3-m^\prime s_2\sigma_2-(k_1+k_2)(v_{1x}s_2+v_{2x}\sigma_2)+(k_2-k_1)v_{1y}s_3.\label{k.p theory}
\eea
which describes the two Dirac cones near $\bar X$ point in SBZ.

\begin{figure}
\includegraphics[width=\columnwidth]{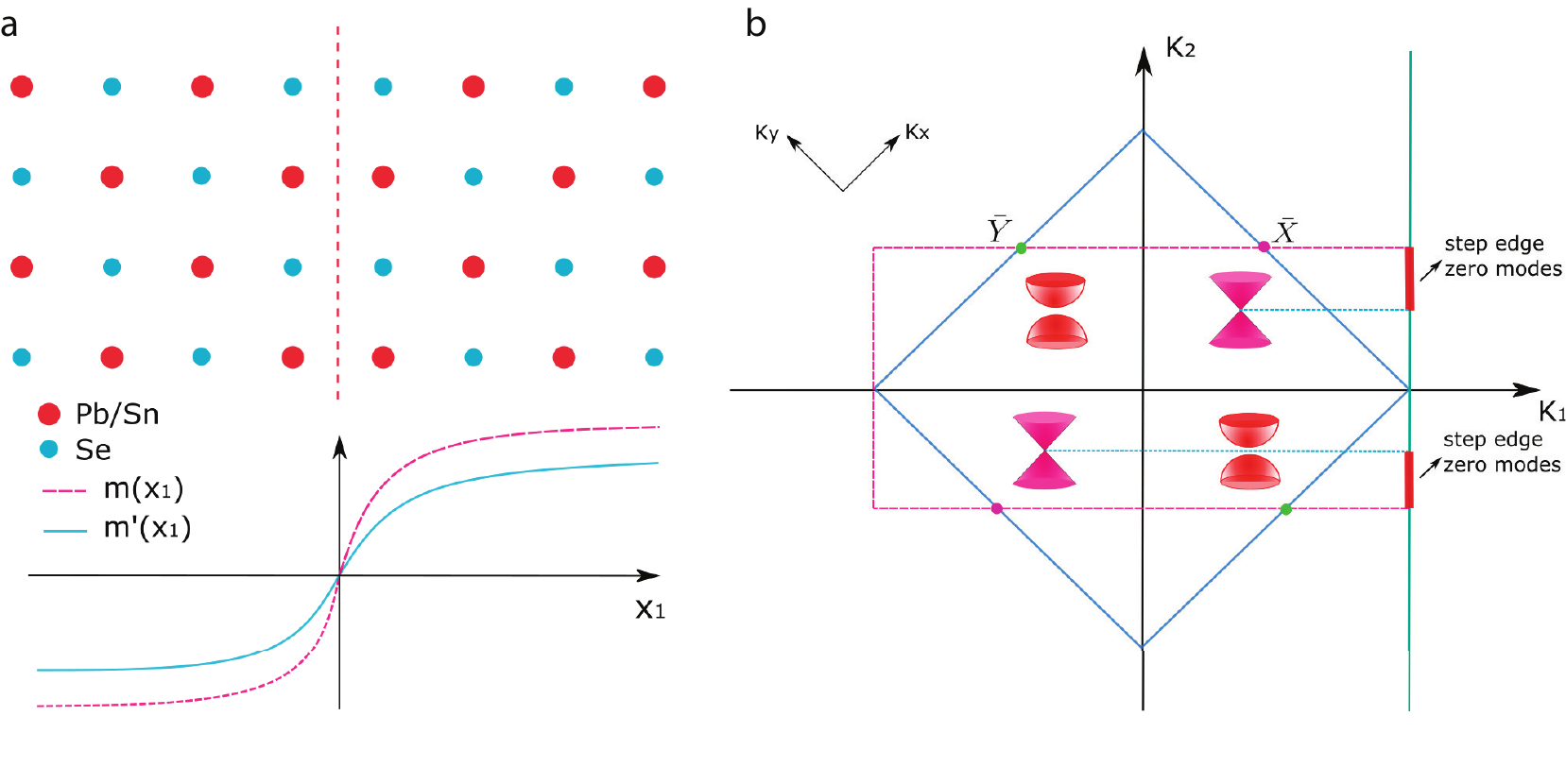}
\caption{Schematic of step edges and surface Brillouin zone. (a) Atomic arrangement, and spatial dependence of parameters $m,m^\prime$ in (\ref{model:odd step edge}) across an odd step edge. (b) Surface Dirac fermions in the [001] surface Brillouin zone (SBZ), and the momentum range of zero modes localized on the odd step edge along (0,1,0) direction.}
\label{fig:c}
\end{figure}

As illustrated in Fig.\ref{fig:c} (b), surface distortions in Pb$_{0.7}$Sn$_{0.3}$Se break the mirror symmetry w.r.t $\hat y$-$\hat z$ plane, opening up a gap ($\sim20$ meV) for the two Dirac cones near $\bar Y$ point\cite{Okada2013}. This leaves the two Dirac fermions in Eq. (\ref{k.p theory}) the only low-energy gapless excitations on the [001] surface. For the convenience of studying step edges along (0,1,0) direction (see Fig.\ref{fig:c}), we have chosen a coordinate system $k_1=(k_x-k_y)/2,~k_2=(k_x+k_y)/2$ for momentum space in model (\ref{k.p theory}) so that the momentum component $k_2$ along the step edge is a good quantum number. As shown by ab initio calcualtions\cite{Wang2013m}, $v_{2x}$ is an order of magnitude smaller than $v_{1x,1y}$ and can be neglected in model (\ref{k.p theory}). Therefore the minimal model for an odd step edge along (0,1,0) direction writes
\bea\label{model:odd step edge}
&\mathcal{H}_{(0,1,0)}(x_1,k_2)=\mathcal{V}(x_1,k_2)+\imth(v_{1x}s_2+v_{1y}s_3)\partial_1,\\
&\mathcal{V}(x_1,k_2)\equiv k_2(v_{1y}s_3-v_{1x}s_2)-m(x_1)\sigma_3-m^\prime(x_1)s_2\sigma_2.\notag
\eea
where $m(x_1)$ and $m^\prime(x_1)$ depend on coordinate $x_1$ across the step edge.

In addition to time reversal symmetry $\bst=\imth s_2\cdot\mathcal{K}$, Hamiltonian (\ref{model:odd step edge}) also exhibits an emergent particle-hole symmetry (PHS) $\bsc=s_1\sigma_2\cdot\mathcal{K}$. At each fixed $k_2$ along the (0,1,0) step edge, the zero modes localized at the edge are classified by the topological indices for $\bsc$- and $\bst$-symmetric 1d insulators\cite{Matsuura2013,Zhao2013,Chiu2016}. They are characterized by an even-integer-valued topological invariant $\nu\in2\mbz$, the winding number of 1d systems in symmetry class AIII\cite{Altland1997,Schnyder2008,Kitaev2009} with chiral symmetry $\chi=\bst\cdot\bsc$. For a (0,1,0) step edge of height $\Delta h=\frac n2a$ ($a$ denotes the lattice constant, see Fig.\ref{fig:1} d-e) on [001] surface of Pb$_{0.7}$Sn$_{0.3}$Se, as we will show below, its topological index is given by $\nu=\pm2$ if $n=$~odd for all momentum $k_2$ between two massless Dirac points around $\bar X$. Similar to the zero-energy flat bands on the zigzag edge of graphene\cite{Nakada1996}, there will be a flat band with $|\nu|$ in-gap states (or $|\nu|/2$ Kramers pairs) at each $k_2$ between the two Dirac cones around $\bar X$ localized at the step edge, as shown in Fig. \ref{fig:c} (b). While odd step edges have a nontrivial topological index $\nu=\pm2$, an even step edge ($n=$~even) generally has a vanishing topological index $\nu=0$.

To show this, we first consider an odd step edge of height $\Delta h=\frac12a$ ($n=1$) described by Hamiltonian (\ref{model:odd step edge}). As depicted in Fig.\ref{fig:c} (a), across an odd step edge the Pb/Sn and Se atoms are switched. Exchange of Pb/Sn and Se orbitals is implemented by operator $\sigma_1$ in model (\ref{model:odd step edge}), leading to the domain wall configuration of $m(x_1)$ and $m^\prime(x_1)$ in Fig.\ref{fig:c}(a):
\bea
\sigma_1\mathcal{H}_{(0,1,0)}(x_1,k_2)\sigma_1=\mathcal{H}_{(0,1,0)}(-x_1,k_2).
\eea
since Pb/Sn and Se orbitals are switched across an odd step edge at $x_1=0$. In particular, the mass domain wall of $m^\prime(x_1)$ at the step edge $x_1=0$ induces a Kramers pair of Jackiw-Rebbi solitons (zero modes) at energy $E=0$:
\bea
\psi_{k_2}(x_1)=e^{\frac{\imth v_-^2k_2x_1-v_{1y}\int_0^{x_1}m^\prime(x)\text{d}x}{v_+^2}} e^{-\frac{\int_0^{x_1}\hat{H}_0(x)\text{d}x}{v_+^2}}\dket{s_1=\sigma_2}.\label{wf:zero mode}
\eea
where we define $v^2_\pm=v_{1y}^2\pm v_{1x}^2$ and
\bea
\hat H_0\equiv \imth v_{1x}m^\prime\sigma_2+2k_2v_{1x}v_{1y}s_1+\imth m(v_{1x}s_2+v_{1y}s_3)\sigma_3.
\eea
The zero-energy Kramers degeneracy in $s_1\sigma_2=1$ subspace cannot be split as long as time reversal is preserved. Moreover, PHS pins it at zero energy (Direc point). It has a one-to-one correspondence to the nontrivial topological index $\nu=2$ of symmetry class AIII associated with this step edge. This explains the observed in-gap \didv~peaks at odd step edges in Fig.\ref{fig:2}c.

As shown in Fig. \ref{fig:c}a, in addition to this step edge at $x_1=0$, there is another type of odd step edge (e.g. the step edge located at $x_1=x_0$ in Fig.\ref{fig:c}a), where the mass $m,m^\prime$ are positive on its l.h.s. and negative on its r.h.s., opposite to the odd step edge at $x_1=0$. Its associated topological index is therefore $\nu=-2$, and there is also a Kramers pair at zero energy on this step edge as protected by time reversal and PHS. The wavefunction of this Kramers pair at $x_1=x_0$ is very similar to (\ref{wf:zero mode}) for the $x_1=0$ odd step edge, except that its low-energy Hilbert space satisfies $s_1\sigma_2=-1$ in contrast to $s_1\sigma_2=+1$ for the $x_1=0$ step edge.

We emphasize that the PHS in the minimal model (\ref{model:odd step edge}) is an emergent symmetry of the surface states at low energy. It is not a microscopic symmetry, for example the small $v_{2x}$ term in surface Hamiltonian (\ref{k.p theory}) in fact weakly violates this PHS. Weak breaking of the PHS will produce a small dispersion for the flat bands at the step edge, and slightly shift it away from the zero energy. Therefore it will slightly shift and broaden the peak in the $dI/dV$ curve at an odd step edge.

With the above understanding of a $\Delta h=\frac12a$ odd step edge, we now consider an even step edge of height $\Delta h=a$. Topologically, such a step edge can be viewed as two odd step edges of $\Delta h=\frac12a$ merging with each other. Therefore the natural low-energy model for such an even step edge is  to consider 2 Kramers pairs (or 4 zero modes) of in-gap flat bands from the 2 constituent odd step edges, and analyze the symmetry-allowed mixing terms between them. First of all, we notice that the topological index of an even step edge is the sum of index for each constituent odd step edge, such as the two step edges in Fig.\ref{fig:c}a, which leads to a trivial topological index $\nu_\text{even}=2-2=0$ for an even step edge. This suggests symmetry-allowed mixing terms exist and will split the two Kramers pairs at an even step edge, pushing them away from zero energy. This explains the observed split peak feature in \didv~maps at even step edges in Fig.\ref{fig:2}d. However, the energy splitting of the two Kramers pairs generally depends on microscopic conditions of the even step edge, and is not a universal quantity. When the splitting is large enough compared to the bulk gap, both Kramers pairs can merge into the bulk states and hence the in-gap peak in \didv~spectra can disappear as reported in \Ref{Sessi2016}.

\begin{figure}
\includegraphics[width=0.6\columnwidth]{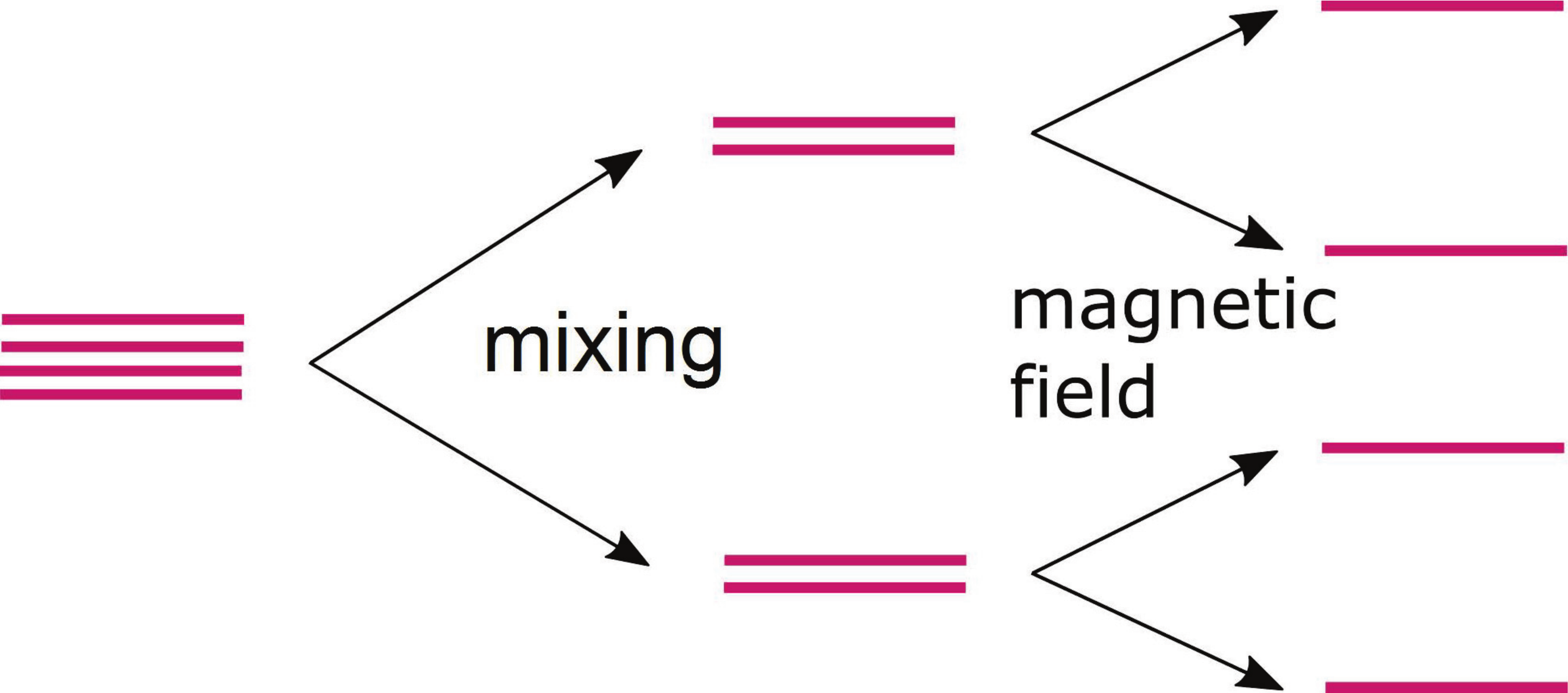}
\caption{Splitting of edge modes. Illustration of how the 4 in-gap flat bands split at an even step edge of height $\Delta h=a$. These bands at even step edges are not topologically protected and can therefore mix and split as seen in the STM data in Fig. 2.}
\label{fig:5}
\end{figure}

Using our theory, we can understand the response of the step edge states to an external magnetic field. For the odd step edge with $\Delta h=\frac12a$, the Kramers pair will split under an external magnetic field in an anisotropic way. As shown in the Appendix, the in-gap zero modes only directly couple to a perpendicular magnetic field at lowest order.  A perpendicular field (along [001] axis) induces a splitting $\sim g\mu_Bh$ proportional to field $h$, and in-plane field will lead to a splitting $\sim h^2/\Delta$ where $\Delta$ stands for the bulk gap. Taking $g=2$ as a rough estimate, a 11 T perpendicular field can induce a splitting of $\sim1$ meV, which is difficult to resolve as was seen in the \didv~curve shown in \Ref{Sessi2016}. For our field of 7.5T, the splitting is of the order of 0.7meV, far less than the peak width and can therefore not be resolved.

At an even step edge with $\Delta h=a$, on the other hand, the two separated Kramers pairs (separated due to mixing) will split into 4 energy levels under a perpendicular magnetic field (Fig.\ref{fig:5}). The theory further predicts that a perpendicular field will move two levels towards the middle of the gap, and move the other two levels up and down towards the bulk bands, as illustrated in Fig.\ref{fig:5}. While the effect is small, it explains the behavior of the splits peaks at the even step edges in magnetic field data at 7.5 T (see Fig.\ref{fig:3} and Fig. 7 in supplement). The linecuts in Fig.2d and Fig.3c were obtained in the same position across the even step edge. One clearly sees a build up of density of states between the two peaks in the magnetic field due to the movement of the peaks as predicted by theory.

\section{Conclusion}

In this paper we establish the topological nature of 1D in-gap states localized at step edges on [001] surface of TCI Pb$_{0.7}$Sn$_{0.3}$Se, both theoretically and experimentally. The emergent particle-hole symmetry (PHS) in the low-energy surface theory gives rise to one Kramers pair of in-gap flat bands at odd step edges of height $\Delta h=(n+\frac12)a$, characterized by a 1d topological winding number $\nu=\pm2$. These step edge states are not spin polarized but rather form a time-reversal-related Kramers pair. Symmetry-allowed mixing between two Kramers pairs lead to the split peak feature at even step edges, consistent with experimental observations in \didv~curves (Fig.\ref{fig:2} d). Under an external magnetic field, each Kramers pair of flat bands can further be split into two levels, which results in a merging of the two split peaks at an even step edge as seen experimentally.

An important point is that although the explicit calculations shown here were carried out for step edges along (0,1,0) direction, as long as Pb/Sn and Se atoms are switched across the step edge, the domain wall configuration of $m$ and $m^\prime$ remains valid and the zero modes follow as Jackiw-Rebbi solitons on the domain wall\cite{Jackiw1976}. Therefore our analysis should apply to step edges along all directions. Finally, we note that our theory is quite general and implies that similar step edge modes can be observed on surface step edges of other topological materials with rock-salt crystal structure.

\section{Methods}

The experiment was performed in an ultra-high vacuum (UHV) system with a base pressure lower than $10^{-10}$ mbar and at a temperature of $\sim$4 K. Scanning tunneling microscopy (STM) and spectroscopy (STS) were used to detect the step edges on the sample surface and to reveal the electronic properties of the 1D conducting channels trapped within them. Single crystal samples of TCI Pb$_{0.7}$Sn$_{0.3}$Se were cleaved at a temperature of $\sim$80 K before being transferred into the STM head. Further details of the theoretical calculations are provided in the supplemental information. \\

\acknowledgements{We thank Paulo Sessi, Ryszard Buczko and Rafal Rechcinski for enlightening discussions, and Ronny Thomale and Titus Neupert for feedback. This work is supported by US Department of Energy under Award Number DE-SC0014335 (STM studies), by NSF Award No. DMR-1610143 (data analysis), by the Center for Emergent Materials, an NSF MRSEC, under award number DMR-1420451 (CYW), and by NSF under award number DMR-1653769 (YML). Part of this work was presented in 2018 APS March Meeting\footnote{https://meetings.aps.org/Meeting/MAR18/Session/C08.11}.}

\bibliographystyle{apsrev4-1}
\bibliography{bibs_career}

\newpage

\begin{widetext}

\begin{center}\large{Supplemental Materials}\end{center}

\appendix

\section{Review of [001] surface Hamiltonian of Pb$_{1-x}$Sn$_x$Se}

The topological crystalline insulator (TCI) Pb$_{1-x}$Sn$_x$Se belongs to space group \#225, $Fm\bar3m$. While the Bravais lattice is expanded by
\bea
{\bf a}_1=a(0,\frac12,\frac12),~~{\bf a}_2=a(\frac12,0,\frac12),~~{\bf a}_3=a(\frac12,\frac12,0)
\eea
the reciprocal lattice is expanded by
\bea
{\bf b}_1=\frac1a(-1,1,1),~~{\bf b}_2=\frac1a(1,-1,1),~~{\bf b}_3=\frac1a(1,1,-1).
\eea
The 8 time reversal invariant momenta (TRIM) are $\vec\Gamma=(0,0,0)$ and
\bea
{\bf X}=\frac{{\bf b}_2+{\bf b}_3}2,~~{\bf Y}=\frac{{\bf b}_1+{\bf b}_3}2,~~{\bf Z}=\frac{{\bf b}_1+{\bf b}_2}2
\eea
and
\bea
{\bf L}_i=\frac12{\bf b}_i,~~i=1,2,3;~~{\bf L}_4=\frac{{\bf b}_1+{\bf b}_2+{\bf b}_3}2.
\eea
This TCI is featured by simultaneous band inversions at the above four symmetry-related TRIM $\{{\bf L}_i|1\leq i\leq4\}$.

The [001] surface of Pb$_{0.7}$Sn$_{0.3}$Se preserves the following symmetries which generate a 2d point group $C_{4v}$:
\bea
&(x,y,z)\overset{C_4^z}\longrightarrow(-y,x,z);\\
&(x,y,z)\overset{M_{[1\bar10]}}\longrightarrow(y,x,z).
\eea
The surface Bravais lattice vectors are
\bea
\bar{\bf a}_1={\bf a}_3=\frac a2(1,1),~~{\bar{\bf a}}_2={\bf a}_1-{\bf a}_2=\frac a2(-1,1).
\eea
with reciprocal lattice vectors
\bea\label{surface bz}
\bar{\bf b}_1=\frac1a(1,1),~~~\bar{\bf b}_2=\frac1a(-1,1).
\eea
There are 4 surface Dirac cones related by $C_4^z$ rotational symmetry, which can be grouped into two pairs:
\bea
{\bf Q}_1=Q(1,1),~~~{\bf Q}_3=-Q(1,1),~~~Q\lesssim\frac{\pi}{a}.
\eea
which preserves mirror symmetry $M_{[1\bar10]}$ w.r.t. $[1\bar10]$ plane; and
\bea
{\bf Q}_2=Q(-1,1),~~~{\bf Q}_4=Q(1,-1).
\eea
which preserves mirror symmetry $M_{[001]}=(C_4^z)^2\cdot M_{[1\bar10]}$ w.r.t. to $[001]$ plane. Note that ${\bf Q}_{1,3}$ are very close to the edge center $\bar X=\frac{\pi}{a}(1,1)$ of the square-shaped surface BZ, while ${\bf Q}_{2,4}$ are close to the other edge center $\bar Y=\frac{\pi}{a}(-1,1)$.

As shown in Ref.\onlinecite{Wang2013m,Zeljkovic2014}, expanding around $\bar X$ point, the effective $k\cdot p$ theory for surface Dirac states at ${\bf Q}_{1,3}$ can be written as
\bea
&\notag h^{\bar X}_{\bf k}=-m\sigma_3-m^\prime s_2\sigma_2-k_x(v_{1x}s_2+v_{2x}\sigma_2+v_{3x}s_2\sigma_3)\\
&-k_y(v_{1y}s_3+v_{2y}s_1\sigma_1+v_{3y}s_3\sigma_3).\label{ham:surface at X}
\eea
in basis of $\{\dket{p_z(\text{Sn}),\uparrow},\dket{p_z(\text{Sn}),\downarrow},\dket{p_x(\text{Se}),\uparrow},\dket{p_x(\text{Se}),\downarrow}\}$. Here $\uparrow/\downarrow$ denotes the spin orientation along $[001]\parallel{\bf a}_3$ crystalline direction (\ie along $\bar\Gamma\bar X$ in the surface BZ). We use $\vec\sigma$ and $\vec s$ to denote Pauli matrices for the orbital and spin indices respectively. We have chosen the following coordinates from surface reciprocal vectors (\ref{surface bz})
\bea
{\bf k}=k_x\bar{\bf b}_1+k_y\bar{\bf b}_2=\frac1a(k_x-k_y,k_x+k_y).
\eea
Under symmetry operations the spinor transforms as
\bea\label{sym:trs}
&\psi_{\bf k}\overset{\bst}\longrightarrow\imth s_2\psi_{-\bf k},\\
\label{sym:mirror X}&\psi_{(k_x,k_y)}\overset{M_{[1\bar10]}}\longrightarrow-\imth s_2\psi_{(k_x,-k_y)},\\
\label{sym:mirror Y}&\psi_{(k_x,k_y)}\overset{M_{[001]}}\longrightarrow\imth\sigma_3s_3\psi_{(-k_x,k_y)}.
\eea
For the $x=0.3$ compound, we have\cite{Wang2013m}
\bea
\notag&m=0.056~\text{eV},~~m^\prime=0.026~\text{eV},\\
\notag&v_{1x}=2.58~\text{eV}\cdot\text{\AA},~~v_{1y}=3.28~\text{eV}\cdot\text{\AA},\\
&v_{2x}=0.32~\text{eV}\cdot\text{\AA},~~v_{3x}=v_{2y}=v_{3y}=0.\label{ham:parameters}
\eea
In our notation, near surface TRIM $\bar X$, the Pauli matrices $\vec s$ correspond to the following physical spin polarization:
\bea\label{spin direction}
s_1\parallel(0,0,1),~~s_2\parallel(1,-1,0),~~s_3\parallel(1,1,0).
\eea

The surface Dirac fermions at ${\bf Q}_{2,4}$ near $\bar Y=\frac{\pi}{a}(-1,1)$ are described similar to (\ref{ham:surface at X}), related by a $C_4^z$ rotation. Experimentally the [001] surface turns out to be structurally distorted, where mirror symmetry $M_{[001]}$ is spontaneously broken. This gaps out the Dirac fermions at ${\bf Q}_{2,4}$ near $\bar Y$, and therefore the only gapless surface states are described by (\ref{ham:surface at X}) and protected by mirror symmetry $M_{[1\bar10]}$ in (\ref{sym:mirror X}).

\section{Step edge states on [001] surface}

\subsection{Setup}

Choosing a different coordinate system for the surface momentum:
\bea
k_1=\frac{k_x-k_y}2,~~~k_2=\frac{k_x+k_y}2
\eea
the surface state Hamiltonian (\ref{ham:surface at X}) can be rewritten as
\bea
h^{\bar X}_{\bf k}=-m\sigma_3-m^\prime s_2\sigma_2-(k_1+k_2)(v_{1x}s_2+v_{2x}\sigma_2)+(k_2-k_1)v_{1y}s_3.\label{ham:surface at X:new basis}
\eea
Now let's consider a step edge along $(0,1,0)$ direction where $k_2$ is still a good quantum number. From real parameters (\ref{ham:parameters}) one can see that $v_{1x},v_{1y}\gg v_{2x}$, and therefore as a minimal model we can neglect $v_{2x}$. It's straightforward to identify that two Dirac points are located at $(k_x,k_y)=\pm(\sqrt{m^2+(m^\prime)^2}/v_{1x},0)$. We can also write down the following Hamiltonian for a $(0,1,0)$ step edge:
\bea\label{ham:step edge}
&\mathcal{H}_{(0,1,0)}=\mathcal{V}(x_1,k_2)+\imth(v_{1x}s_2+v_{1y}s_3)\partial_1,\\
&\mathcal{V}(x_1)\equiv-m(x_1)\sigma_3-m^\prime(x_1)s_2\sigma_2+k_2(v_{1y}s_3-v_{1x}s_2).\notag
\eea

\subsection{Topological classification of step edge states}

In addition to time reversal symmetry (\ref{sym:trs}), the above minimal model (\ref{ham:step edge}) also preserves a particle-hole symmetry $\mathcal{C}$
\bea\label{sym:phs}
&\psi_{\bf k}\overset{\mathcal{C}=s_1\sigma_2}\longrightarrow s_1\sigma_2\psi_{-{\bf k}}^\ast,\\
&\mathcal{C}\cdot\mathcal{H}_{(0,1,0)}(k_2)\cdot\mathcal{C}=-\mathcal{H}^T_{(0,1,0)}(-k_2)
\eea

Therefore the total symmetry of the minimal model (\ref{ham:step edge}) is generated by $U(1)$ charge symmetry generated by $\mathcal{Q}=e^{\imth\frac\pi2\hat F}$ ($\hat F$ denotes the total fermion number), time reversal symmetry $\bst$ and particle-hole symmetry $\mathcal{C}$, satisfying the following commutation relations:
\bea
\mathcal{C}^2=(-1)^{\hat F},~~[\bst,\mathcal{C}]=\{\mathcal{C},\mathcal{Q}\}=\{\bst,\mathcal{Q}\}=0.
\eea
The step edge bound state is classified by the extension problem of complex Clifford algebra:
\bea
\{\gamma_1,\bst\mathcal{C}\}\times\mathcal{Q}\rightarrow\{\gamma_1,\bst\mathcal{C},\gamma_0\}\times\mathcal{Q}
\eea
where $\gamma_{1,0}$ are Dirac matrices describing one side of the step edge, as a 1d system with a fixed $k_2$ since only the combination $\bst\mathcal{C}$ of time-reversal and particle-hole operations preserves momentum $k_2$. This leads to a classification of
\bea
\nu\in\pi_0(\mathcal{C}_2)=\mbz
\eea
characterized by an integer-valued winding number $\nu$ in symmetry class AIII (since $\bst\mathcal{C}$ can be regarded as the chiral symmetry). For the 1d system at $k_2=0$, both time reversal $\bst$ and particle-hole symmetry $\mathcal{C}$ are present, also leading to an integer classification of $\pi_0(R_{3-1+2})=\mbz$. Due to the time reversal symmetry at $k_2=0$ and Kramers theorem, the winding number of the 1d Hamiltonian at a general fixed $k_2\approx0$ (near $\bar X$) must be an even integer
\bea
\nu=2\times(\#~\text{of Kramers pairs})\in2\mbz
\eea

Notice that particle-hole symmetry $\mathcal{C}$ in (\ref{sym:phs}) is only a special property of minimal model (\ref{ham:step edge}). A generic surface Hamiltonian (\ref{ham:surface at X}) will not have this particle-hole symmetry $\mathcal{C}$, \eg the small $v_{2x}$ term in (\ref{ham:surface at X:new basis}) breaks the symmetry (\ref{sym:phs}). With only time reversal symmetry the 1d system has a trivial classification (\ie symmetry class AII), hence breaking of particle-hole symmetry will generally split the zero-energy bound states at the step edge away from zero energy. The splitting will be roughly proportional to the size of particle-hole symmetry breaking in the surface states.

\subsection{Localized zero modes at odd step edge}

Here we solve the minimal model (\ref{model:odd step edge}) explicitly for the odd step edge\cite{Sessi2016} illustrated in FIG. \ref{fig:c}, to demonstrate the existence of zero modes for small $k_2\approx0$ localized at the step edge.

Eigenstates of step edge Hamiltonian (\ref{model:odd step edge}) satisfy the following Schrodinger equation:
\bea
\imth(v_{1x}s_2+v_{1y}s_3)\partial_1\psi_{E,k_2}=\big(E-\mathcal{V}(x_1,k_2)\big)\psi_{E,k_2}
\eea


For zero modes with $E=0$, we have
\bea
\notag&(v_{1x}^2+v_{1y}^2)\partial_1\psi_{0,k_2}=-H_0(x_1,k_2)\psi_{0,k_2}\\
&+[\imth(v_{1y}^2-v_{1x}^2)k_2-v_{1y}m^\prime(x_1)s_1\sigma_2]\psi_{0,k_2},\\
&H_0(x)\equiv \imth v_{1x}m^\prime(x)\sigma_2+2k_2v_{1x}v_{1y}s_1+\imth (v_{1x}s_2+v_{1y}s_3)\sigma_3 m(x),~~~~~[\hat H_0,s_1\sigma_2]=0\notag
\eea
As illustrated in FIG. \ref{fig:c}, across an odd step edge the Pb/Sn and Se atoms are switched, as if the whole Hamiltonian (\ref{model:odd step edge}) is acted by $\sigma_1$ operator which exchanges Pb/Sn and Se orbitals. This leads to the domain wall configuration of $m(x_1)$ and $m^\prime(x_1)$ as shown in FIG. \ref{fig:c}. In the presense of this mass domain wall at the odd step edge, the above Schrodinger equation for step edges has the following solution:
\bea\label{zero mode}
&\psi_{0,k_2}(x_1)=e^{\imth\frac{v_-^2}{v_+^2}k_2x_1-\frac{v_{1y}}{v_+^2}\int_0^{x_1}m^\prime(x)\text{d}x}\cdot\mathcal{X}e^{-\frac1{v_+^2}\int_0^{x_1}\hat{H}_0(x,k_2)\text{d}x}\psi_{0,k_2}(0),\\
&s_1\sigma_2\psi_{0,k_2}(0)=\psi_{0,k_2}(0).
\eea
where $\mathcal{X}$ denotes ordering w.r.t. coordinate $x$ for the integral and
\bea
v_\pm^2=v_{1y}^2\pm v_{1x}^2.
\eea
Notice that the eigenvalues of non-Hermitian operator $\hat H_0(x_1,k_2)$ are given by
\bea
&\lambda_\pm(x_1,k_2)=\pm\sqrt{(2k_2v_{1y}+\imth m^\prime)^2v_{1x}^2-m^2v_+^2}.
\eea
As long as the following condition is satified
\bea
|\text{Re}\lambda_\pm|=\frac{2k_2v_{1x}^2m^\prime v_{1y}}{\sqrt{(mv_+)^2+(v_{1x}m^\prime)^2}}+O(|k_2|^2)<v_{1y}m^\prime
\eea
the zero mode wavefunction (\ref{zero mode}) is always localized around the step edge at $x_1=0$. This indicates a flat band of zero modes at all small $k_2\approx0$ at the odd step edge, as illustrated in FIG. \ref{fig:c}(b).

Notice that
\bea
\bst s_1\sigma_2\bst^{-1}=s_1\sigma_2
\eea
therefore the zero modes described in (\ref{zero mode}) form one Kramers pair, which cannot split due to time reversal symmetry. However, applying an external magnetic field can split the two zero modes.

As shown in Fig.\ref{fig:c}(a), aside from the step edge at $x_1=0$, there is another type of odd step edge at $x_1=x_0$. The mass functions $m(x_1)$ and $m^\prime(x_1)$ again change sign across the step edge at $x_1=x_0$, but from positive to negative this time, in contrast to the odd step edge at $x_1=0$. A calculation completely in parallel to the $x_1=0$ step edge can be carried out: the corresponding zero mode wavefunction also has the form of Eq. (\ref{zero mode}), except that the subspace of zero modes at the $x_1=x_0$ step edge satisfies
\bea
s_1\sigma_2=-1
\eea
instead of $+1$ for the step edge at $x_1=0$. These two types of odd step edges are characterized by topological index $\nu=+2$ at $x_1=0$ and $\nu=-2$ at $x_1=x_0$ respectively. When we bring the two step edges closer to each other by decreasing their distance $x_0$, these two pairs of Kramers doublets will be mixed and can be gap out each other without breaking the PHS, since together they have a total topological index of $\nu_\text{tot}=+2-2=0$.

\subsection{Splitting of zero modes by the magnetic field}

Below we quantitatively compute the splitting of the Kramers pair localized at the (0,1,0) odd step edge in the presence of an external Zeeman field. Note that both zero modes of the Kramers pair satisfy
\bea
s_1\sigma_2=\pm1
\eea
depending on the type of odd step edge. Therefore the effective spin-$1/2$ Pauli matrices acting within the zero modes' subspace are
\bea
\vec\tau\equiv(s_1,s_2\sigma_3,s_3\sigma_3)=\pm(\sigma_2,-s_3\sigma_1,s_2\sigma_1).
\eea
Let's label the Kramers pair of zero modes in (\ref{zero mode}) as $\dket{\tau_z=+1}\equiv\dket{\uparrow}$ and $\dket{\tau_z=-1}\equiv\dket{\downarrow}$. Their matrix elements under an external magnetic field
\bea
\delta\hat H=-\vec{h}\cdot\vec{s}
\eea
are given by
\bea
\mathcal{P}_0\delta\hat H\mathcal{P}_0=-h_1\tau_x
\eea
where $\mathcal{P}_0$ denotes the projection operator into the zero-modes subspace. Note that only out-of-plane magnetic field along (0,0,1) direction will split the two zero modes by $h_1$. Meanwhile any in-plane field $h_{2,3}$ will only mix these two zero modes with other states, causing a splitting $\sim(h_{2,3})^2/\Delta$ where $\Delta$ is the energy difference between the zero modes and other high-energy states.

\subsection{Even step edges}

The simplest model for an even step edge is to consider two odd step edges (such as the two step edges in Fig.\ref{fig:c}a) very close to each other, so that the zero modes at each odd step edge can mix with each other and split. As their distance decreases, the two odd step edges merge into an even step edge. Therefore we start from the low-energy degrees of freedom at the two odd step edges, i.e. the two Kramers doublets to describe the low-energy physics of an even step edge.

Here we use Dirac matrices $\vec\mu$ for the flavor index of the two odd step edges, and $\vec\tau$ for the Kramers doublet index at each step edge. By appropriately choosing the basis, the time reversal symmetry in the Hilbert space of four zero modes at the even step edge can be written as
\bea
\bst = \imth\tau_2\cdot\mathcal{K},
\eea
while particle-hole symmetry is implemented by
\bea
\bsc=\mu_3\cdot\mathcal{K}
\eea
The possible mixing terms preserving both time reversal and particle-hole symmetries have the following form
\bea
\delta H_\text{even}=\delta_1\mu_1+\delta_2\tau_2\mu_2.
\eea
and they will split the 4 zero modes into two Kramers pairs at an even step edge. The energy splitting $|\delta|\equiv\sqrt{\delta_1^2+\delta_2^2}$ between the two Kramers pairs generally depends on the microscopic condition of the even step edge, and is not a universal quantity.

Similar to an odd step edge, in the lowest order the zero modes at an even step edge only couples to an external magnetic field along $s_1$ i.e. (0,0,1) direction, and generally the Hamiltonian of an even step edge under a magnetic field can be written as
\bea
\delta\hat H_\text{even}=-h_1\tau_1+\delta_1\mu_1+\delta_2\tau_2\mu_2.
\eea
Its spectrum is $E=\pm \sqrt{(h_1\pm\delta_1)^2+(\delta_2)^2}$.
We find that the magnetic field can further split these two Kramers pairs. With a small magnetic field $h_1\ll |\delta|$, the gap $2\sqrt{(h_1-\delta_1)^2+(\delta_2)^2}$ between two energy levels in the middle scales as $\sim 2|\delta|-2\delta_1h_1/|\delta|$, and decreases with an increasing magnetic field. When the field reaches $h_1=\delta_1$, the middle gap reaches its minimum $|\delta|_\text{min}=|\delta_2|$. With a large field  $h_1\gg |\delta|$, the gap scales linearly with field as $\sim 2h_1$.

\newpage

\subsection{Linecuts across the even step at 0T and 7.5T}

\begin{figure}[h]
\includegraphics[width=0.5\columnwidth]{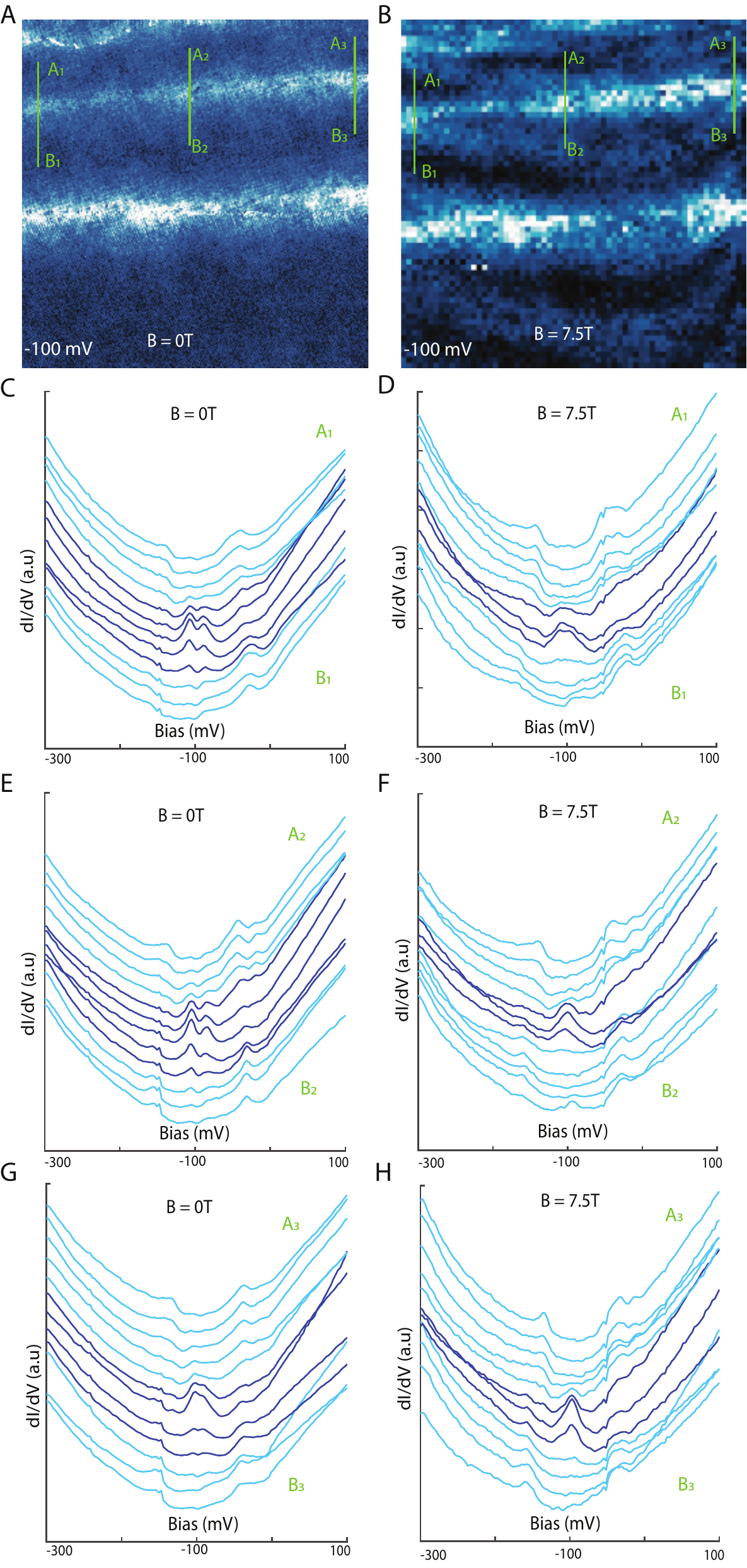}
\caption{\textbf{Linecuts across the even step.} \textbf{(A)} dI/dV map at 0T, -100 mV. The green lines indicate the positions of the linecuts. \textbf{(B)} dI/dV map at 7.5T. The area and positions of the linecuts are the same of (A).\textbf{(C)}, \textbf{(E)}, and \textbf{(G)}  linecuts at 0T along the positions shown in (A). \textbf{(D)}, \textbf{(F)}, and \textbf{(H)} linecuts at 7.5T along the positions shown in (B) which are the same as (A)}
\label{fig:Sup}
\end{figure}

\end{widetext}

%
%
%

\end{document}